\documentclass[12pt]{article}

\usepackage{graphicx,cite}

\newcommand{\be}{\begin{equation}}
\newcommand{\ee}{\end{equation}}
\newcommand{\bea}{\begin{eqnarray}}
\newcommand{\eea}{\end{eqnarray}}
\newcommand{\ba}{\begin{array}}
\newcommand{\ea}{\end{array}}

\begin{document}


\begin{flushright}
  HD-THEP-04-28\\
  hep-ph/0407224
\end{flushright}

\vspace{\baselineskip}

\begin{center}
\textbf{\LARGE Optimal observables\\[0.3em]
        and phase-space ambiguities\\}
\vspace{4\baselineskip}
{\sc O.~Nachtmann\footnote{email:
O.Nachtmann@thphys.uni-heidelberg.de} and F.~Nagel\footnote{email:
F.Nagel@thphys.uni-heidelberg.de}}\\
\vspace{1\baselineskip}
\textit{Institut f\"ur Theoretische Physik, Philosophenweg 16, D-69120
Heidelberg, Germany}\\
\vspace{2\baselineskip}
\textbf{Abstract}\\
\vspace{1\baselineskip}
\parbox{0.9\textwidth}{Optimal observables are known to lead to minimal
  statistical errors on parameters for a given normalised event distribution
  of a physics reaction.  Thereby all statistical correlations are taken into
  account.  Therefore, on the one hand they are a useful tool to extract
  values on a set of parameters from measured data.  On the other hand one can
  calculate the minimal constraints on these parameters achievable by any
  data-analysis method for the specific reaction.  In case the final states
  can be reconstructed without ambiguities optimal observables have a
  particularly simple form.  We give explicit formulae for the optimal
  observables for generic reactions in case of ambiguities in the
  reconstruction of the final state and for general parameterisation of the
  final-state phase space.}
\end{center}
\vspace{\baselineskip}

\pagebreak


\section{Introduction}
\label{sec-intro}

At a future \mbox{$e^+ e^-$} linear collider~(LC) like
TESLA~\cite{Richard:2001qm} or CLIC~\cite{Ellis:1998wx}, apart from the
possible discovery of new particles, electroweak precision measurements of
various observables will be an important task.  In this way effects due to new
physics at a scale far beyond the produced c.m.~energy may be detected.  The
\mbox{$e^+ e^-$}, the \mbox{$\gamma \gamma$}, the \mbox{$e^- \gamma$}, the
\mbox{$e^- e^-$} and the Giga-$Z$ mode can reveal complementary aspects.  For
instance, consider the three-gauge-boson vertices \mbox{$\gamma WW$}
and~\mbox{$ZWW$}, which are highly restricted in the Standard Model~(SM), see
for instance~\cite{Nachtmann:ta}.  In a form-factor approach with the most
general Lorentz-invariant parameterisation these vertices are described by 28
real parameters if one allows also for imaginary parts~\cite{Hagiwara:1986vm}.
In many extensions of the~SM one obtains deviations of these
triple-gauge-boson couplings~(TGCs) from their SM~values, for an overview of
the literature see e.g.~Sect.~1 of~\cite{Diehl:2002nj}.  At present the
experimental constraints on these couplings from~LEP are rather weak and many
couplings have yet to be
measured~\cite{Abreu:1999vv,Acciarri:2000kf,Heister:2001qt}.  At a future~LC
all~TGCs can be measured with much higher precision,
see~\cite{Diehl:1993br,Diehl:1997ft,Menges:2001gg} and references therein.
Both longitudinal and transverse beam polarisation are considered to be
feasible~\cite{Moortgat-Pick:2002gb} and will be advantageous in many cases,
see for example~\cite{Moortgat-Pick:2001kg}.  For most TGCs longitudinal
polarisation is found to be the right choice~\cite{Diehl:2002nj}, whereas one
coupling can only be determined with transverse
polarisation~\cite{Diehl:2003qz}.  In a gauge invariant effective-Lagrangian
approach for the gauge-boson sector new-physics effects can be parameterised
by ten anomalous couplings~\cite{Buchmuller:1985jz,Nachtmann:2004ug}.  In such
a framework, after spontaneous symmetry breaking also gauge-boson-fermion
interactions and gauge-boson masses are modified by the anomalous couplings.
Therefore constraints can be derived from observables measured at LEP1, SLC
and~LEP2.  Also in this case the couplings can be determined with much higher
precision at a future~LC~\cite{Nachtmann:2004ug}.

In all cases it is important to know how sensitive the event distribution of a
certain reaction is to a given set of parameters, independent of the method
that will be used in the future to analyse the data.  Moreover statistical
correlations should be taken into account, unless they are very small, in
order to provide realistic results.  Frequently one is dealing with a
situation where the normalised event distribution is---at least
approximately---a linear function of the parameters to be measured.  The
method of optimal observables~\cite{Atwood:1991ka,Diehl:1993br} is most
directly applied for the case of a strictly linear function.  In most parts of
this paper we assume that the event distributions are to good approximation
linear functions of the parameters to be measured.  By means of optimal
observables one can then compute the maximum constraints on this set of
parameters for a given reaction and given event number while taking into
account all statistical correlations.  Apart from being a useful tool for
theorists to estimate the sensitivity of a reaction to a set of parameters
optimal observables have been used in experimental analyses to extract the
parameters, for instance TGCs~\cite{Abreu:1999vv} or $CP$~violating parameters
in~\mbox{$e^+ e^- \rightarrow \tau^+ \tau^-$}~\cite{Ackerstaff:1996gy}.  But
also methods to handle non-linear functions of the parameters to be measured
have been devised~\cite{Diehl:1997ft} and successfully applied in experimental
analyses, see~e.g.~\cite{bib:dham,Heister:2001qt}.

In case of many parameters with correlated errors, the sensitivity to
different directions in parameter space is often not easy to survey.  Here
optimal observables have the advantage that discrete-symmetry properties of
the differential cross section can be exploited in order to eliminate
correlations between couplings.  For example the~TGCs are classified into four
symmetry groups, and couplings from different symmetry groups can be measured
without correlations when optimal observables are used~\cite{Diehl:1993br}.
However within each symmetry group couplings are in general correlated.
In~\cite{Diehl:1997ft} it is shown that the optimal observables are unique up
to linear transformations.

Here we construct explicitly the optimal observables for generic ambiguities
in the reconstruction of the final state.  Such ambiguities can be either
continuous, that is the number of measured variables is smaller than the
number of variables required to specify the final state, or discrete, that is
a reconstructed event can originate from two or more final states.

This work is organised as follows: In Sect.~\ref{sec-opt} we resume the
properties of optimal observables as given in~\cite{Diehl:1993br}.  In
Sect.~\ref{sec-ooamb} the optimal observables for generic ambiguities in the
reconstruction of the final state are discussed.  In Sect.~\ref{sec-nonlin} we
recall that the method can be applied iteratively to determine parameters on
which the differential cross section depends non-linearly.  We present our
conclusions in Sect.~\ref{sec-concl}.


\section{Optimal observables}
\label{sec-opt}

In this section we give a resum\'{e} of the definition and properties of
optimal observables.  As it is convenient to have an illustrative example in
mind we shall discuss the problem to measure anomalous contributions to the
differential cross sections for the reactions \mbox{$e^+e^- \rightarrow WW$}
and \mbox{$\gamma \gamma \rightarrow WW$}.  But we emphasise that our
considerations are neither restricted to anomalous couplings nor to particular
reactions.  The optimal-observables method can be applied to any reaction
where the differential cross section depends on a certain number of small
parameters, which we generically denote by~$h_i$ and which are to be
estimated.  Furthermore, the method can be generalised to the case where the
parameters to extract are not necessarily small~\cite{Diehl:1997ft}.  Consider
now the reaction \mbox{$e^+e^- \rightarrow WW$} or \mbox{$\gamma \gamma
\rightarrow WW$} and let us assume that we describe it in the framework of
the~SM with the addition of small real anomalous constants~$h_i$.

In an experiment one measures the differential cross section
\be
\label{eq-diffpol}
S(\phi) \equiv {\rm d}\sigma / {\rm d}\phi ,
\ee
where $\phi$ denotes the set of all measured phase-space variables.  For
instance, the fully differential cross section of the process \mbox{$e^+ e^-
\rightarrow WW$} without transverse beam polarisation depends on five angles,
which are in this case specified by~$\phi$, see~\cite{Diehl:2002nj}.  In the
same way the spin-averaged fully differential cross section of \mbox{$\gamma
\gamma \rightarrow WW$} with fixed photon energies depends on five angles,
see~\cite{photon-collider}.  In case of \mbox{$e^+ e^- \rightarrow WW$} as
treated within a form-factor approach in~\cite{Diehl:2002nj} the anomalous
parameters~$h_i$ are the 28 anomalous~TGCs that parameterise deviations at the
\mbox{$\gamma WW$} and \mbox{$ZWW$} vertices from the~SM.  In case of
\mbox{$\gamma \gamma \rightarrow WW$} as treated within an
effective-Lagrangian approach in~\cite{photon-collider} the~$h_i$ are the ten
coefficients of certain dimension-six operators.  Here non-zero anomalous
couplings do not only lead to anomalous three- and four-gauge-boson couplings
but also to deviations of the gauge-boson-fermion couplings and of the
gauge-boson masses from their SM~values.  We distinguish between the
information from the total cross section \mbox{$\sigma = \int {\rm d}\sigma$}
and from the normalised distribution \mbox{$S/\sigma$} of the events.  Here we
only investigate how well anomalous couplings can be extracted from the
latter.  It is possible to obtain constraints on these parameters also from
the measurement of~$\sigma$, see Sect.~3.1 of~\cite{Diehl:1997ft}.  Those
considerations remain unaffected in the presence of ambiguities in the
reconstruction of the final state.

Expanding $S$ in the anomalous couplings one can write
\be
\label{eq-distri}
S(\phi) = S_0(\phi) + \sum_i S_{1i}(\phi) \, h_i \; + \; O(h^2),
\ee
where \mbox{$S_0(\phi)$} is the cross section in the~SM and the
\mbox{$S_{1i}(\phi)$} give the first-order modifications due to the anomalous
couplings.  We assume \mbox{$S_0(\phi)$} and \mbox{$S_{1i}(\phi)$} to be
calculated from theory.  Note that the variables~$\phi$ need not specify the
final state completely.  In the analyses~\cite{Diehl:2002nj,Diehl:2003qz}
of~\mbox{$e^+e^- \rightarrow WW$} with one~$W$ decaying leptonically and the
other one into two hadronic jets it was assumed that the jet charges cannot be
identified, which results in a two-fold ambiguity.  In such cases
\mbox{$S(\phi)$} is not the fully differential cross section---in our example
it is the sum over two final states.  In a case with ambiguities of this or
more involved kind it is often not straightforward to calculate
\mbox{$S_0(\phi)$} and~\mbox{$S_{1i}(\phi)$}.  But these quantities must be
known explicitly in order to construct the optimal observables.  This problem
is our main concern in this paper and will be addressed in the following
section.

We now give a short resum\'{e} of the optimal-observables method.  One way to
extract the parameters~$h_i$ from the measured distribution~(\ref{eq-distri})
is to look for a suitable set of observables~\mbox{$\mathcal{O}_i (\phi)$}
whose expectation values
\be
\label{eq-exdef}
E[\mathcal{O}_i] = \frac{1}{\sigma} \int \!{\rm d}\phi\, S(\phi) \,
\mathcal{O}_i (\phi)
\ee
are sensitive to the dependence of~\mbox{$S(\phi)$} on the couplings~$h_i$.
To first order in the anomalous couplings we have
\be
\label{eq-exexp}
E[\mathcal{O}_i] = E_0[\mathcal{O}_i] + \sum_j c_{ij} h_j \; + \; O(h^2),
\ee
with
\bea
\label{eq-exsmobs}
E_0[\mathcal{O}_i] & = & \frac{1}{\sigma_0} \int \!{\rm d}\phi\, S_0 (\phi) \,
\mathcal{O}_i (\phi) \;, \\[.5ex]
\label{eq-cij}
c_{ij}     & = & \frac{1}{\sigma_0} \int \!{\rm d}\phi\, \mathcal{O}_i (\phi)
\, S_{1j} (\phi) - \frac{\sigma_{1j}}{\sigma_0^{2}} \int \!{\rm d}\phi\, S_0
(\phi)\, \mathcal{O}_i (\phi)\;,\\[.5ex]
\label{eq-tcsm}
\sigma_0 & = & \int \!{\rm d}\phi\, S_0 (\phi) \;,\\[.5ex] 
\label{eq-tcf}
\sigma_{1j} & = & \int \!{\rm d}\phi\, S_{1j} (\phi) \;.
\eea
Here $E_0[\mathcal{O}_i]$ is the expectation value for zero anomalous
couplings, and $c_{ij}$ gives the sensitivity of~$E[\mathcal{O}_i]$ to~$h_j$.
Solving~(\ref{eq-exexp}) for the set of the~$h_j$ we get estimators for the
anomalous couplings, whose covariance matrix is given by
\be
\label{eq-vh}
V(h) = \frac{1}{N}\, c^{-1} V(\mathcal{O})\, (c^{-1})^{\rm T},
\ee
where we use matrix notation.  Here $N$ is the number of events, and
\be
\label{eq-covo}
V(\mathcal{O})_{ij} = \frac{1}{\sigma_0}
             \int \!{\rm d}\phi\, S_0 (\phi) \, \mathcal{O}_i (\phi) \,
             \mathcal{O}_j (\phi) -
             E_0[\mathcal{O}_i] \, E_0[\mathcal{O}_j]\;\;+\;\;O(h)
\ee
is the covariance matrix of the observables, which we have expanded around its
value in the~SM.  As observables we choose
\be
\label{eq-defob}
\mathcal{O}_i (\phi) = \frac{S_{1i}(\phi)}{S_0(\phi)}.
\ee
{}From~(\ref{eq-cij}) and~(\ref{eq-covo}) one obtains for this specific choice
of observables
\be
\label{eq-ccov}
V(\mathcal{O}) = c \; + \; O(h),
\ee
and therefore
\be
\label{eq-geo}
V(h) = \frac{1}{N }\, c^{-1} \; + \; O(h).
\ee
{}From~(\ref{eq-ccov}) we see that for the observables~(\ref{eq-defob}) $c$ is
a symmetric matrix because~\mbox{$V(\mathcal{O})$} is symmetric.  The
observables~(\ref{eq-defob}) are ``optimal'' in the sense that for $h_i
\rightarrow 0$ the errors~(\ref{eq-geo}) on the couplings are as small as they
can be for a given probability distribution, see~\cite{Diehl:1993br}.  For
details on this so-called Rao-Cram\'er-Fr\'echet bound, see for
example~\cite{Cramer:1958}.  Apart from being useful for actual experimental
analyses, the observables~(\ref{eq-defob}) thus provide insight into the
sensitivity that is at best attainable by {\em any} method, given a certain
process and specified experimental conditions.  In case of one parameter this
type of observable was first proposed in~\cite{Atwood:1991ka}, the
generalisation to several parameters was made in~\cite{Diehl:1993br}.
Moreover, it has been shown that optimal observables are unique up to a linear
reparameterisation~\cite{Diehl:1997ft}.  We further note that phase-space
cuts, as well as detector efficiency and acceptance have no influence on the
observables being ``optimal'' in the above sense, since their effects drop out
in the ratio~(\ref{eq-defob}).  This is not the case for detector-resolution
effects, but the observables~(\ref{eq-defob}) are still close to optimal if
such effects do not significantly distort the differential
distributions~$S_{1i}$ and~$S_0$ (or tend to cancel in their ratio).  To the
extent that they are taken into account in the data analysis, none of these
experimental effects will bias the estimators.

When a set of events is calculated with a Monte-Carlo generator and the events
are then reconstructed like in an experimental analysis the optimal
observables~(\ref{eq-defob}) are obtained directly in the measured
variables~$\phi$.  However for theoretical studies, i.e.\ to estimate the
sensitivity of a certain reaction to anomalous couplings, and also for
experimental analyses often the analytical expressions of the optimal
observables~(\ref{eq-defob}) are required.

Frequently there are ambiguities in phase space, that is to one value of the
measured kinematic variables~$\phi$ there correspond two or more distinct
final states (discrete ambiguities) or a bunch of final states (continuous
ambiguities).  The calculation of~\mbox{$S_0 (\phi)$}, \mbox{$S_{1i} (\phi)$}
in~(\ref{eq-distri}) and in particular of the observables~\mbox{$\mathcal{O}_i
(\phi)$} in~(\ref{eq-defob}) has then to be done with some care as will be
shown below.


\section{Phase-space ambiguities}
\label{sec-ooamb}

In principle there are plenty of possibilities to parameterise a final state
in a reaction uniquely, for instance the usage of angles or Cartesian
coordinates, different choices of reference frames etc.  In an experiment one
may either be able to specify a final state of an event uniquely or only with
certain ambiguities.  Here we discuss in detail the case of discrete
ambiguities, that is for each event one only knows that it belongs to a group
of two, three or more final states.  We also mention how to handle continuous
ambiguities, i.e.\ the case in which the number of measured variables is
smaller than the number of variables required to specify the final state.  An
example of a discrete ambiguity is the two-fold one of the semileptonic final
states in \mbox{$e^+ e^- \rightarrow WW$} or in \mbox{$\gamma \gamma
\rightarrow WW$} with fixed c.m.~energy of the two-photon system.  Here one
usually assumes that the two hadronic jets cannot be associated unambiguously
to the quark and antiquark.  Another more involved one occurs in the reaction
\mbox{$\gamma \gamma \rightarrow WW$} when the photons each obey a Compton
spectrum.  Here, in addition to the ambiguity above, another two-fold one
arises from the reconstruction of the neutrino momentum.  This case is
considered in~\cite{photon-collider}.  The optimal
observables~(\ref{eq-defob}) are now to be calculated in the presence of such
ambiguities.

We start from a particular set of phase-space variables~$\chi$ that specify
the final state uniquely.  The differential cross section in terms of these
variables we denote by
\be
T(\chi) \equiv {\rm d}\sigma / {\rm d}\chi.
\ee
The cross section for another choice of variables~$\phi$, that may lead to the
above ambiguities, is then given by
\be
\label{eq-tdef}
S (\phi) = \int {\rm d}\chi \; \delta (F(\chi) - \phi) \, T(\chi).
\ee
The function~$F$ expressing the relation of~$\phi$ to~$\chi$ may take the same
value for different values of~$\chi$, that is for a given~$\phi$ the equation
\be
\label{eq-solf}
F(\chi) = \phi
\ee
may have several solutions~\mbox{$\chi_k \equiv \chi_k(\phi)$} with \mbox{$k =
  1,2,\ldots$}.  In general, the number of solutions to~(\ref{eq-solf}) may
  vary with~$\phi$.  If $\phi$ are the coordinates that can be measured of an
  event~$\chi$, the set of final states~$\chi_k$ consists of $\chi$ itself as
  well as all final states that cannot be distinguished from~$\chi$ by a
  measurement of~$\phi$.  Notice that in~(\ref{eq-tdef}) we have assumed that
  the number of components of the vectors~$\phi$ and~$\chi$ are the same.  In
  general there can be also continuous ambiguities, that is the fully
  differential cross section is specified by a larger number of variables than
  those that can actually be measured.  However this does not lead to any
  further complicacies in the context of optimal observables.  In fact, if the
  fully differential cross section is~\mbox{$\tilde{T} (\chi, \xi)$} where the
  final-state variables~$\xi$ cannot be measured, the generalisation
  of~(\ref{eq-tdef}) is
\be
\label{eq-tdef2}
S (\phi) = \int {\rm d}\chi {\rm d}\xi \; \delta (F(\chi) - \phi) \,
\tilde{T}(\chi, \xi).
\ee
If we define
\be
T(\chi) \equiv \int {\rm d}\xi \; \tilde{T}(\chi, \xi)
\ee
we again obtain~(\ref{eq-tdef}).  We can thus apply all formulae in the
remainder of this section also in case of discrete {\it plus} continuous
ambiguities.  We remark that our analysis also works if one or more of the
phase-space variables take discrete values as is the case for instance for
spin indices.  For these variables integrals have to be substituted by sums
and $\delta$-distributions by Kronecker symbols.

An integration and summation over part of the phase-space variables as
in~(\ref{eq-tdef2}) is, of course, performed when one considers inclusive
cross sections.  Thus our discussion covers also this case.  Clearly, then the
normalisation integral \mbox{$\int {\rm d}\phi \; S(\phi)$} gives the cross
section times the corresponding multiplicity and in our formulae~$\sigma$ has
to be read in this way.

Coming back to~(\ref{eq-tdef}) we have
\be
\label{eq-t2}
S (\phi) = \sum_k |J_k|^{-1} \; T(\chi_k(\phi))
\ee
where
\be
\label{eq-jacob}
J_k \equiv \det \frac{\partial F}{\partial \chi} (\chi_k(\phi))
\ee
is the Jacobian determinant taken at point~$\chi_k$.  If $F$ is invertible,
there is only one term in the sum for all~$\phi$ and~(\ref{eq-t2}) simplifies
to
\be
S(\phi) = \left|\frac{\partial F}{\partial
    \chi}\left(F^{-1}(\phi)\right)\right|^{-1} T\left(F^{-1}(\phi)\right).
\ee
We expand the differential cross section:
\be
T(\chi) = T_0(\chi) + \sum_i T_{1i}(\chi) \, h_i \;\; + \;\; O(h^2).
\ee
It follows
\be
\label{eq-sphi}
S(\phi) = S_0(\phi) + \sum_i S_{1i}(\phi) \, h_i \;\; + \;\; O(h^2),
\ee
where
\bea
\label{eq-t0xi}
S_0(\phi) & = & \sum_k |J_k|^{-1} \; T_0(\chi_k(\phi)),\\
\label{eq-t1xi}
S_{1i}(\phi) & = & \sum_k |J_k|^{-1} \; T_{1i}(\chi_k(\phi)).
\eea
Note, again, that the number of terms in the sums~(\ref{eq-t0xi})
and~(\ref{eq-t1xi}) can vary with~$\phi$.  If $\phi$---but not
necessarily~$\chi$---are coordinates that can be measured we have to define
the optimal observables from the expansion of~\mbox{$S(\phi)$}
in~(\ref{eq-sphi}):
\be
\label{eq-defobamb}
\mathcal{O}_i(\phi) = \frac{S_{1i}(\phi)}{S_0(\phi)}.
\ee
In the specific case where $F$ is invertible going from~$\chi$ to~$\phi$ is a
mere change of coordinates and we obtain the same optimal observables using
either set of variables:
\be
\label{eq-obspec}
\mathcal{O}_i (\phi) = \left. \frac{T_{1i}\left(\chi \right)}{T_0\left(\chi
    \right)}\right|_{\chi \, = \, F^{-1}(\phi)}.
\ee
If there are ambiguities in the reconstruction but if we have the same
Jacobian~\mbox{$J \equiv J_k$} for all~$k$ (which may nevertheless depend
on~$\phi$), $J$ cancels in the numerator and denominator of the
observables~(\ref{eq-defobamb}):
\be
\label{eq-sumobspec}
\mathcal{O}_i(\phi) = \frac{\sum_k
T_{1i}\left(\chi_k(\phi)\right)}{\sum_k T_0\left(\chi_k(\phi)\right)}.
\ee
This is the case e.g.\ for the reaction \mbox{$e^+e^- \rightarrow WW$}, where
one $W$~boson decays into a quark-antiquark pair and the other one into a
lepton pair, if the charges of the two jets in the final state cannot be
identified, see~\cite{Diehl:1993br}.  If this is not the case we must use the
general expressions~(\ref{eq-t0xi}) to~(\ref{eq-defobamb}).

\begin{figure}
\centering
\includegraphics[totalheight=6cm]{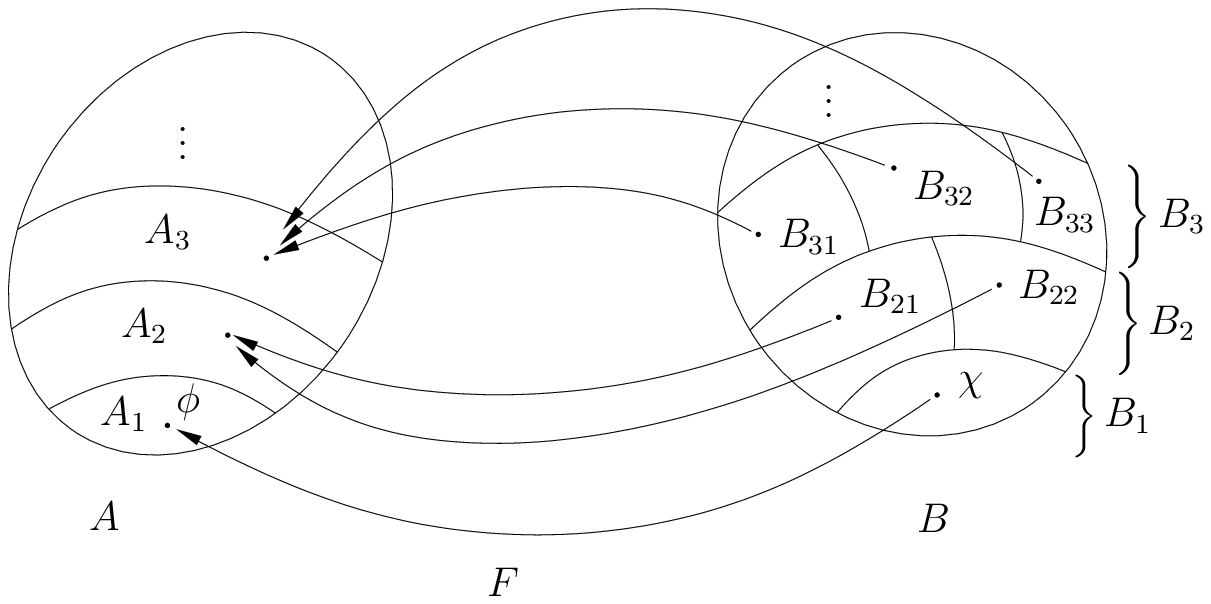}
\caption{\protect{\label{fig:sets}}Definition of integration areas.}
\end{figure}

The covariance matrix of the observables~(\ref{eq-defobamb}) is now
\be
\label{eq-covoamb}
V(\mathcal{O})_{ij} = \frac{1}{\sigma_0}
             \int_A \!{\rm d}\phi \; S_0 (\phi) \, \mathcal{O}_i (\phi) \,
             \mathcal{O}_j (\phi) -
             \frac{\sigma_{1i} \sigma_{1j}}{\sigma_0^2}\;\;+\;\;O(h),
\ee
where
\bea
\sigma_0 & \equiv & \int_A \!{\rm d}\phi \; S_0(\phi) = \int_B \!{\rm d}\chi
\;T_0(\chi),\\
\sigma_{1i} & \equiv & \int_A \!{\rm d}\phi \; S_{1i}(\phi) = \int_B \!{\rm
  d}\chi \; T_{1i}(\chi),
\eea
and the full kinematically allowed integration regions in the
coordinates~$\phi$ and~$\chi$ are denoted by~$A$ and~$B$, respectively.  The
integrals~$\sigma_0$ and~$\sigma_{1i}$ can be performed in either coordinates.
Using~$\chi$ no knowledge about ambiguities in the reconstruction is
necessary.  The first term in the expression of~\mbox{$V(\mathcal{O})$} needs
special care.  We divide the integration region~$A$ into parts~$A_n$
with~\mbox{$n = 1,2,\ldots$}, such that for~\mbox{$\phi \in A_n$} there are
$n$ solutions~$\chi_k$ to~(\ref{eq-solf}).  The domains of~$B$ corresponding
to the~$A_n$ we denote by~$B_n$, see Fig.~\ref{fig:sets}.
We subdivide $B_n$ into $n$ appropriate regions $B_{nk}$ with \mbox{$k =
1,2,\ldots,n$}, such that \mbox{$\chi_k \in B_{nk}$}.  This subdivision is
certainly not unique.  We have
\bea
H_{ij} & \equiv & \int_A \!{\rm d}\phi \;S_0(\phi) \,
\mathcal{O}_i(\phi) \, \mathcal{O}_j(\phi)\\
 & = & \int_A \!{\rm d}\phi \;\frac{S_{1i}(\phi)S_{1j}(\phi)}{S_0(\phi)}
 \nonumber\\
 & = & \sum_{n \geq 1} \int_{A_n} \!{\rm d}\phi
 \;\frac{S_{1i}(\phi)S_{1j}(\phi)}{S_0(\phi)}. \nonumber  
\eea
Expressed in terms of integrals over~$\chi$ we get
\be
\label{eq-vint1}
H_{ij} = \int_{B_1} \!{\rm d}\chi \;\frac{T_{1i}(\chi) \,
T_{1j}(\chi)}{T_0(\chi)} \; + \; \sum_{n \geq 2} \int_{B_{np_n}} \!{\rm
d}\chi \;|J(\chi)| \frac{S_{1i}\left(F(\chi)\right)
S_{1j}\left(F(\chi)\right)}{S_0\left(F(\chi)\right)},
\ee
and also
\be
\label{eq-vint2}
H_{ij} = \int_{B_1} \!{\rm d}\chi \;\frac{T_{1i}(\chi) \,
T_{1j}(\chi)}{T_0(\chi)} \; + \; \sum_{n \geq 2} \frac{1}{n} \int_{B_n}
\!{\rm d}\chi \;|J(\chi)| \frac{S_{1i}\left(F(\chi)\right)
S_{1j}\left(F(\chi)\right)}{S_0\left(F(\chi)\right)},
\ee
where we have from~(\ref{eq-t0xi}), (\ref{eq-t1xi}) for \mbox{$\chi \in B_n$}
\bea
\label{eq-t0}
S_0 \left(F(\chi)\right) & = & \sum_k^n
\left|J\Big(\chi_k(F(\chi))\Big)\right|^{-1}
T_0 \Big(\chi_k(F(\chi))\Big),\\
\label{eq-t1i}
S_{1i} \left(F(\chi)\right) & = & \sum_k^n
\left|J\Big(\chi_k(F(\chi))\Big)\right|^{-1}
T_{1i}\Big(\chi_k(F(\chi))\Big),\\[.2cm]
J(\chi) & = & \det \frac{\partial F}{\partial \chi} (\chi).
\eea
One of the values \mbox{$\chi_k(F(\chi))$} in~(\ref{eq-t0}), (\ref{eq-t1i})
is, of course, identical to~$\chi$.  In~(\ref{eq-vint1}) one can choose for
each~$n$ any natural number~$p_n$ with \mbox{$1 \leq p_n \leq n$}.  These
choices correspond to different parameterisations of the integration
regions~$A_n$ but leave the integrals unchanged.  Therefore one may sum over
all possible choices and divide by~$n$, which leads to~(\ref{eq-vint2}).  The
quantities~\mbox{$H_{ij}$} may be calculated either in the
form~(\ref{eq-vint1}) or~(\ref{eq-vint2}).  Notice that the
form~(\ref{eq-vint2}) has the advantage that one only has to know where in the
integration region for~$\chi$ there are how many solutions to~(\ref{eq-solf}),
but one does not have to specify $B_{n1}$, $B_{n2}$, etc.  In certain cases
the integrals for \mbox{$n \geq 2$} in~(\ref{eq-vint1}) or~(\ref{eq-vint2})
may be simplified.  For example let $A'_n$ with \mbox{$n \geq 2$} be the part
of~$A_n$ where the Jacobians \mbox{$J(\chi_k(\phi))$} are the same for
all~$k$.  The Jacobian in this region may nevertheless depend on~$\phi$.  The
region of~$A_n$ where they are not the same for all~$k$ we call~$A^{\prime
\prime}_n$.  The corresponding regions of $B_n$ are denoted by~$B'_n$
and~$B^{\prime \prime}_n$.  We write the integrals in~(\ref{eq-vint2}) as
\be
\int_{B_n} \!{\rm d}\chi = \int_{B'_n} \!{\rm d}\chi +
\int_{B^{\prime \prime}_n} \!{\rm d}\chi.
\ee
Then, in the integrals over $B'_n$ the Jacobian cancels and we obtain the
following expression for the integral in the covariance
matrix~(\ref{eq-covoamb}):
\bea
H_{ij} & = &  \int_{B_1} \!{\rm d}\chi \;\frac{T_{1i}(\chi) \,
  T_{1j}(\chi)}{T_0(\chi)}\\[.8ex]
&&\hspace{-.5cm} {} + \;\sum_{n \geq 2} \frac{1}{n} \int_{B'_n}
\!{\rm d}\chi \; \frac{\sum_k^n T_{1i}\Big(\chi_k(F(\chi))\Big)
 \sum_l^n T_{1j}\Big(\chi_l(F(\chi))\Big)}{\sum_m^n
T_0\Big(\chi_m(F(\chi))\Big)}\nonumber\\[.8ex]
&&\hspace{-.5cm} {} + \;\sum_{n \geq 2} \frac{1}{n} \int_{B^{\prime
\prime}_n} \!{\rm d}\chi \;|J(\chi)| \frac{S_{1i}\left(F(\chi)\right)
S_{1j}\left(F(\chi)\right)}{S_0\left(F(\chi)\right)}\nonumber
\eea
with \mbox{$S_0 (F(\chi))$} and~\mbox{$S_{1i} (F(\chi))$} as in~(\ref{eq-t0})
and~(\ref{eq-t1i}), respectively.


\section{Iterative analysis in the non-linear case}
\label{sec-nonlin}

In this section we recall briefly that the use of optimal observables is not
restricted to a phase-space distribution depending only linearly on
parameters~$h_i$ which are to be estimated.  In other words, the higher-order
terms in~$h_i$ in~(\ref{eq-distri}) can be handled.  This has been discussed
extensively in~\cite{Diehl:1997ft}.  We recall here only one practical
procedure one can follow.

Suppose we have a theoretical expression for the differential cross
section~(\ref{eq-distri}) which can be expanded in the~$h_i$:
\be
\label{eq-ssec}
S(\phi) = S_0(\phi) + \sum_i S_{1i}(\phi) \, h_i + \sum_{ij} S_{2ij} (\phi) \,
h_i h_j \; + \; \ldots.
\ee
With given data an estimate of the~$h_i$ has to be made.  The procedure
proposed in~\cite{Diehl:1997ft} is then as follows.  In the first step the
terms of second and higher order are neglected and one follows the procedure
of estimating the~$h_i$ by the optimal observables~(\ref{eq-defob}).  Suppose
that this gives as estimates for the parameters the values~$\tilde{h}_i$.  In
the second step one sets
\be
h_i = \tilde{h}_i + h'_i
\ee
and substitutes this for~\mbox{$S(\phi)$} in~(\ref{eq-ssec}).  This gives
\be
\label{eq-stilde}
S(\phi) = \tilde{S}_0(\phi) + \sum_i \tilde{S}_{1i}(\phi) \, h'_i + \sum_{ij}
\tilde{S}_{2ij}(\phi) \, h'_i h'_j \; + \; \ldots,
\ee
where
\bea
\tilde{S_0}(\phi) & = & S_0(\phi) + \sum_i S_{1i}(\phi) \, \tilde{h}_i +
\sum_{ij} S_{2ij}(\phi) \, \tilde{h}_i \tilde{h}_j \; + \; \ldots,\\  
\tilde{S}_{1i}(\phi) & = & S_{1i}(\phi) + 2 \sum_j S_{2ij}(\phi) \, \tilde{h}_j
\; + \; \ldots,
\eea
and so on.

Now one applies the optimal-observables method to estimate the~$h'_i$,
neglecting terms of second and higher order in the~$h'_i$
in~(\ref{eq-stilde}).  The new optimal observables are
\be
\tilde{\mathcal{O}}_i(\phi) = \frac{\tilde{S}_{1i}(\phi)}{\ba{c} \\[-.45cm]
  \tilde{S}_0(\phi)\ea}.
\ee
Let~$\tilde{h}'_i$ be the estimates for these parameters obtained in this way.
The improved estimate for the original parameters is then~\mbox{$\tilde{h}_i +
\tilde{h}'_i$}.

This procedure can be iterated.  It was tested in~\cite{bib:dham} in a
Monte-Carlo study for the analysis of TGCs at~LEP2.  Parameters~$h_i$---not
necessarily small---were assumed and Monte-Carlo data generated according to
the corresponding distribution~(\ref{eq-ssec}) which for this case contained
linear and quadratic terms in the~$h_i$.  The non-linear optimal-observables
analysis was then performed as mentioned above.  It turned out that after a
few, typically three, iterations the input values for the~$h_i$ were obtained
back within their correct statistical errors.

Clearly, the phase-space ambiguities are to be treated in exactly the way
discussed in Sect.~\ref{sec-ooamb} also for the case of a non-linear analysis.


\section{Conclusions}
\label{sec-concl}

For electroweak precision measurements at a future~LC it will be important to
check the validity of the~SM (or perhaps another theory) with the highest
possible precision.  To this end optimal observables are a convenient means
because parameters can be determined with minimal errors as allowed by a
theorem from mathematical statistics without neglecting correlations between
any of them.  Such observables have for instance been applied to the reaction
\mbox{$e^+ e^- \rightarrow WW$} in~\cite{Diehl:1993br,Diehl:1997ft} with a
form-factor approach to the \mbox{$\gamma WW$} and \mbox{$ZWW$}~vertices.
Effects of beam polarisation to the same process were analysed
in~\cite{Diehl:2002nj,Diehl:2003qz}.  However in the mentioned studies the
final state was assumed to be known either exactly or up to a two-fold
ambiguity of a very simple type.  It was possible to add the respective terms
of the differential cross section in order to construct the optimal
observables.  In this paper we have discussed the calculation of optimal
observables for the case of an arbitrary reaction with generic ambiguities in
the reconstruction of the final states.  This is the case e.g.\ in the
reaction \mbox{$\gamma \gamma \rightarrow WW$} where the photons are not
monochromatic but have a Compton-energy spectrum~\cite{photon-collider}.  In
the most general case the expressions for the optimal observables and the
covariance matrix are somewhat complicated, basically because they contain the
Jacobian determinant of the parameter transformation.  However simplifications
occur in various special cases, for example if the Jacobian determinant is a
constant in phase space and therefore cancels in certain ratios and integrals.
Using our results one is able to study the best statistically achievable
sensitivity to a set of parameters in a reaction given a certain event number,
if there are ambiguities in the reconstruction of the final state.  Such
studies give important information on the capabilities of future machines and
on how to choose the experimental settings like polarisations in an optimal
way for physics studies.  Apart from that optimal observables are an ideal
tool for the analysis of experimental data.  In either case one often needs
explicit analytical expressions of the optimal observables.  To this end we
have collected here the necessary formulae for the case when the final state
cannot be fully reconstructed from the measured variables.


\section*{Acknowledgements}

The authors are grateful to M.~Diehl for reading a draft version of the
manuscript and for good suggestions and to M.~Pospischil for useful
discussions.  This work was supported by the German Bundesministerium f\"ur
Bildung und Forschung, BMBF project no.~05HT4VHA/0, and the Deutsche
Forschungsgemeinschaft through the Graduiertenkolleg ``Physikalische Systeme
mit vielen Freiheitsgraden''.


\end{document}